\documentclass[twocolumn,showpacs,preprintnumbers,amsmath,amssymb]{revtex4}

\usepackage{graphicx}
\usepackage{dcolumn}
\usepackage{bm}

\begin{document}

\title{Spin-polarization effects in the processes of synchrotron radiation
and electron-positron pair production by a photon in a magnetic field}

\author{O.P. Novak}
\email{novak-o-p@ukr.net}

\author{R.I. Kholodov}
\email{kholodov@yahoo.com}

\affiliation{National Academy of Sciences of Ukraine, Institute of Applied
Physics, 58, Petropavlivska
Street, 40030, Sumy, Ukraine}

\date{\today}

\begin{abstract}Spin and polarization effects and correlations between them
in the processes of pair production by a photon and synchrotron radiation
in a magnetic field are considered. Expressions for the probabilities of the
processes  with arbitrary polarizations of the particles  are obtained.
These expressions are analyzed in detail in both the Lowest Landau Levels and
ultrarelativistic approximations.
\end{abstract}

\maketitle
%==============================================================================

\section{Introduction}
\label{intro} The study of the quantum-electrodynamic processes involving
photons and electrons in strong external electromagnetic fields is still
topical from both an experimental and theoretical point of view,
despite an extensive literature on this subject. The first relativistic
theory of the processes of synchrotron radiation and pair production in a
magnetic field was investigated in the works \cite{Klepikov}-%
\cite{FIAN} in the approximation of ultrarelativistic motion of
particles. The results of these works are included in the monographs of Sokolov
and Ternov \cite{Sokolov1}, \cite{Sokolov2}. The operator
method for solving this problem was applied by Baier and Katkov in the
quasiclassic ultrarelativistic case \cite{Baier1,Baier2}. Recently, there has
appeared the work of these authors \cite{BaierAr}, where the operator method
was used to study the process of electron-positron pair production by a
photon, when the particles are located at low-energy Landau levels.
In the Ref.~\cite{Herold}, synchrotron radiation of electron-positron plasmas
has been studied and Landau level splitting due to interaction with photon
field is taken into account.
In the Ref.~\cite{Preece}, the influence of electron spins on radiation
probability for the first 500 levels has been considered.
Reference \cite{Pavlov} is devoted to the studying of radiative width of
cyclotron line and level splitting due to interaction with QED-vacuum. We also
mention Refs.~\cite{Schwinger}-\cite{Semionova}, where the processes of photon
radiation and pair production was considered for the case of polarized
particles. It should be noted that correlations  between spin and
polarization effects in these processes have not been studied in detail yet.

The purpose of this paper is theoretical research of
spin and polarization effects and their correlation in the processes of
synchrotron radiation and pair production by a
photon in a strong magnetic field. We use general expressions for
probability of the processes when spin projections of the particles and
photon polarization are arbitrary. In this paper the Stokes parameters
are used to define photon polarization.
Expressions for the probabilities are analyzed in the ultraquantum (Lowest
Landau Levels) and the ultrarelativistic approximations that are most
important for experimental applications.
In these approximations, simple analytical expressions depending on both
particles' spins and photon polarization  were obtained. Thus, it turned out
to be possible to carry out analysis of spin and polarization effects and
correlations between them.

Carrying out of corresponding experiments implies usage of magnetic fields that
are comparable with the critical Schwinger one
$B_c = m^2c^3 / e\hbar \approx 4.41 \cdot 10^{13}$~G
and are not feasible in terrestrial laboratories.
The greatest constant field obtained is about $100$~T
\cite{Maglab} and the greatest pulse field is $\sim 10^6$~G \cite{Arzamas}.
Nevertheless, we should point out the possibility to obtain a strong magnetic
field on QED length of about $10^{-11}$~cm \cite{PAST}. In heavy ion collisions
\cite{Darmstadt} (Darmstadt, GSI) Coulomb fields compensate and the magnetic
field can reach a strength of $10^{12}$~G in the region between ions if the
impact parameter is about $10^{-11}$~cm. In principle, QED processes can be
observed in this region.

The investigation of the QED processes keeps actuality and great importance
in view of existence of strong magnetic fields around neutron stars
\cite{Harding6}. Particularly, cyclotron lines have been found in the radiation
of X-ray pulsars. These lines correspond to the cyclotron radiation
(absorption) of electrons that occupy the lowest Landau levels. A lot of works
are devoted to the investigation of these lines \cite{Bussard}-\cite{Harding4}.
References \cite{Baring1}-\cite{Thompson} considering the pair production
process and its applications to the pulsars are worth mentioning too.

It should be noted that the astrophysical modeling of pulsars implies
that radiation is emitted by unpolarized particles.
However, electron-positron plasma in the magnetosphere of a pulsar is mostly
created by the pair production process. Consequently, electron spins in Landau
levels are not equally populated. In Section \ref{app} we compare
transition rates for the cases of polarized and unpolarized particles.

%==============================================================================

\section{Spin-polarization effects in the process of synchrotron radiation}
\label{polefmb}
In a uniform, homogenous magnetic field $B'=B/B_c$ energy levels of electrons
are
\begin{equation}
E_n=\sqrt{p_z^2+m^2+2lB'm^2}
\end{equation}
where $l$ is the principal quantum number ($l=0,1,2 ...$), and $p_z$ is the
momentum component parallel to the field  (we will use natural units, where
$\hbar=c=1$ throughout). In each Landau state, the electron may have spin-up
($s=+1$) or spin-down ($s=-1$) along the field direction, except in the ground
state, where only the spin-down state is allowed.

The procedure of obtaining the probabilities of first-order processes is
well known and we omit the corresponding calculations. The resulting
expressions for probabilities of the processes of synchrotron radiation and
pair production that depend on particles' spins as well as on photon
polarization are given in the appendixes. Now let us proceed directly to
the analysis of the probabilities.

%------------------------------------------------------------------------------
\subsection{Ultraquantum approximation}
In the Lowest Landau Levels (LLL) approximation, intensity
distribution of synchrotron radiation is presented
by the expressions (\ref{eq53})--(\ref{eq56}).

The "no spin-flip" processes
have the greatest probability because they have the lowest power of the
small parameter ${B' \ll 1}$. Moreover, due to the condition $l>l'$ the
radiation probability is maximal for the process with particle spins
directed against the field. The energetically unfavorable process with
particles polarizations $s=1$, $s'=-1$ has the smallest probability.
Process probability decreases as $(B')^{l-l'}$ if the difference $l-l'$
increases therefore the transition $l \to l-1$ is the most probable.

The "no spin-flip" processes have identical dependence of
probability on the Stokes parameters. Probability of the energetically
favorable spin-flip process (\ref{eq55}) differs in the sign of the Stokes
parameter $Q$, therefore radiation has opposite linear
polarization for spin-up -- spin-down transitions.

Let us consider two opposite cases of linear photon polarization. If the
polarization vector is perpendicular to the vector of a magnetic field
$\vec B$ then the Stokes parameter $Q$ equals $-1$, $Q=-1$.
Hereafter we will call it perpendicular photon polarization. If the
polarization vector belongs to the same plane as the wave vector $\vec k$
and the vector $\vec B$ then equation $Q=1$ is true and
such polarization is called parallel polarization.

In the case of parallel polarization ($Q = 1$) the probability of radiation
equals zero in the direction perpendicular to the field ($\theta = \pi/2$)
for the processes without flip of spin. Probability of the spin-down--spin-up
transition (\ref{eq56}) is minimal in this direction. Probability of
the other spin-flip process is a slowly varying function of the polar angle.
In the case of perpendicular polarization ($Q=-1$) radiation is absent in the
direction $\theta = \pi/2$ for the spin-flip processes, but probabilities of
the transitions without flip of spin depend on the polar angle weakly.

Radiation has circular polarization in the direction along the magnetic field,
since the probability does not depend on the parameter of linear polarization
$Q$ if the condition $\cos\theta = \pm 1$ is fulfilled. If the photon has
right circular polarization ($Q = 1$), radiation probability is maximal
in the direction along the magnetic field ($\theta = 0$) and equals zero in
the direction against the field. In the case of left polarization ($Q= - 1$),
the situation is reversed.

One can see that the polarization of radiation is the same as in the case of
classical motion of an electron. As follows from the above, substantial
spin-polarization correlation takes place. The shape of the angular
distribution of radiation probability and its representative values are
determined by the values of photon polarization and spin projections of
the particles.

\begin{figure}
\begin{center}
\resizebox{\columnwidth}{!}
{  \includegraphics{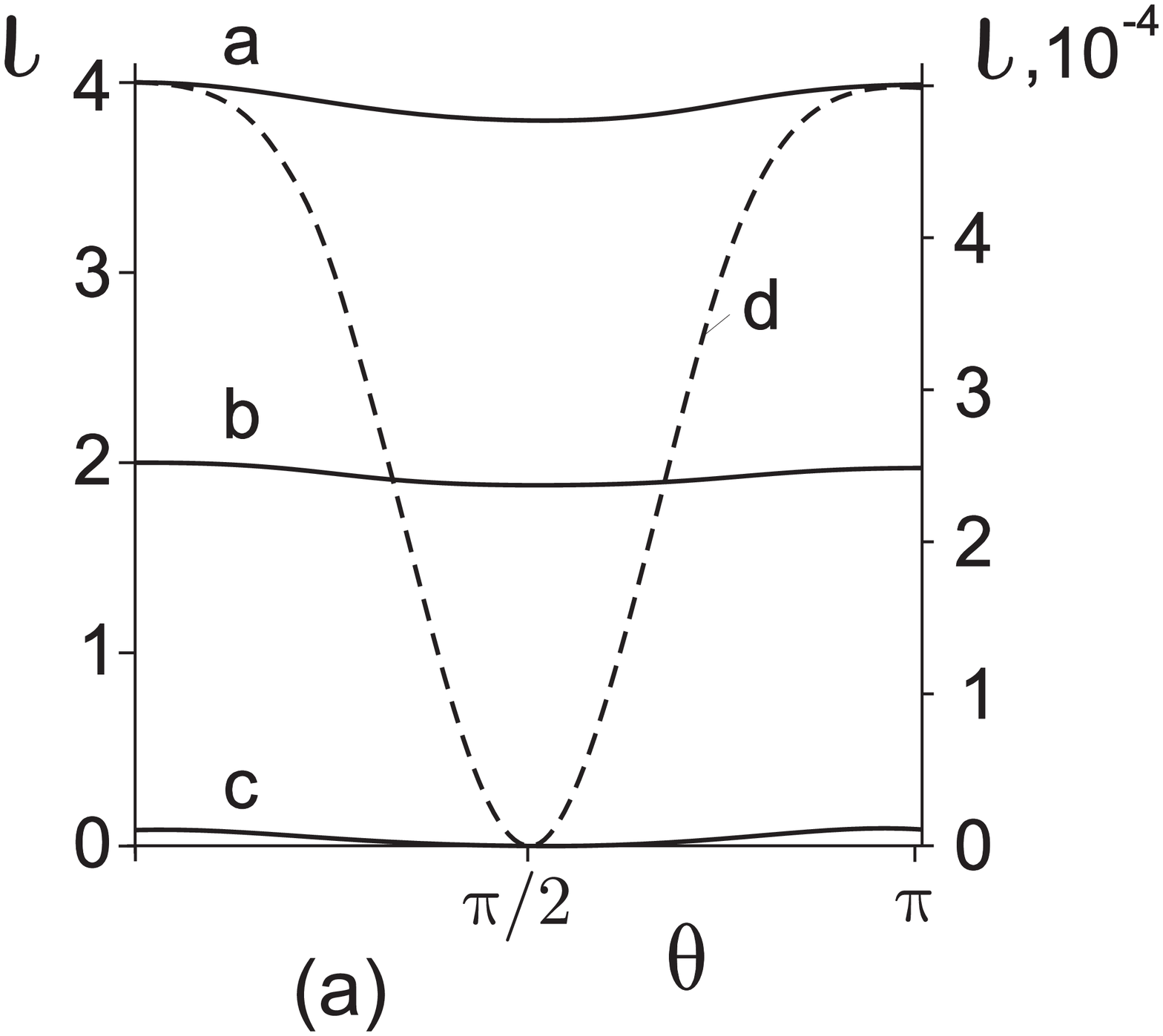} }
\resizebox{\columnwidth}{!}
{  \includegraphics{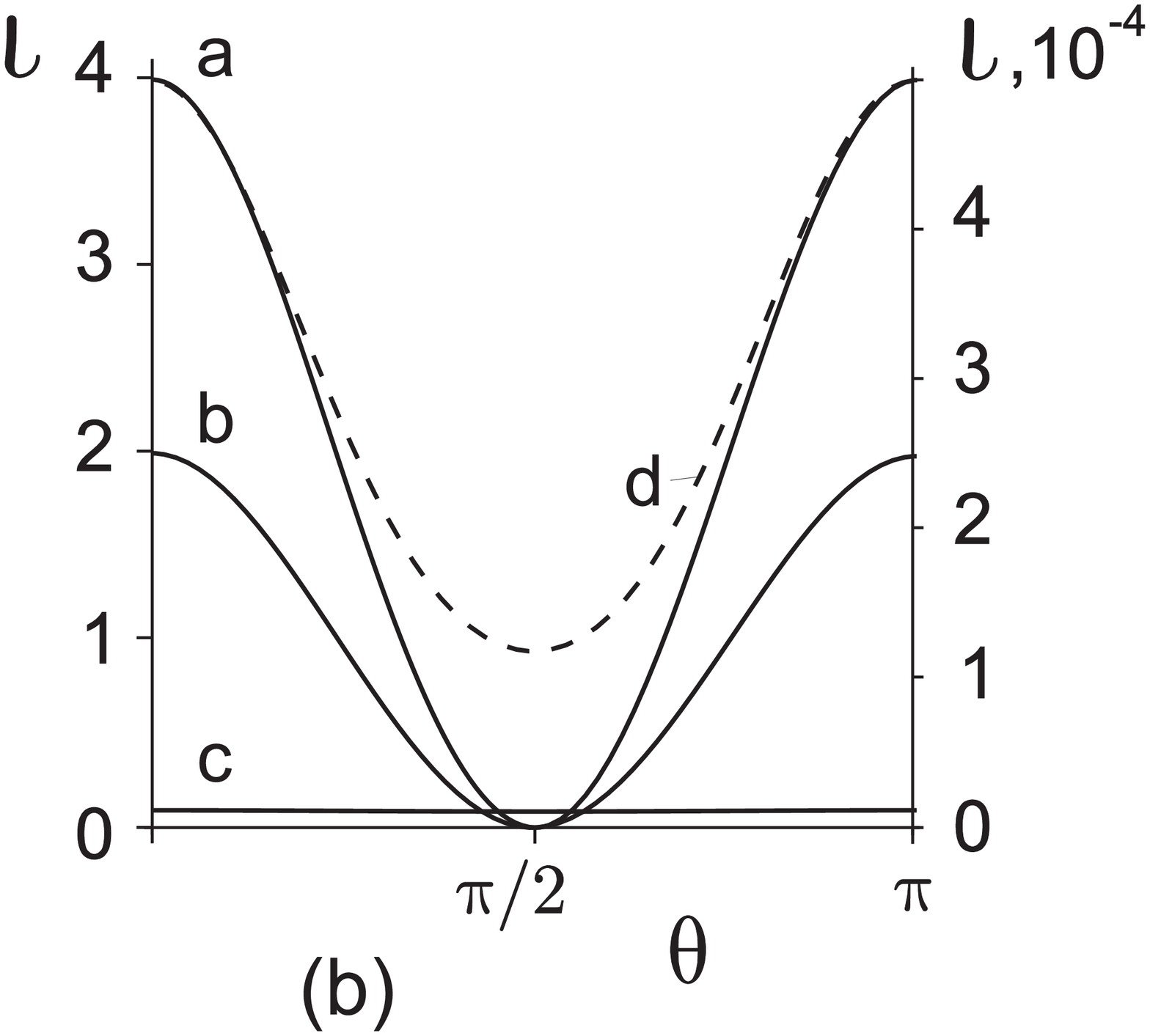} }
\end{center}
\caption{
Angular distribution of the radiation intensity of perpendicular (a) and parallel
(b) linear polarization. The value of the field is $B' = 0.1$.
Here: (a) spin-down--spin-down transition; (b) spin-up--spin-up transition;
(c) $s =1$, $s ' = - 1$; (d) $s = - 1$, $s ' = 1$ (broken line).
The right scale relates to the dashed line (d).}
\label{fig:3}
\end{figure}

The angular distributions of intensity in relative units for linear
polarization of the radiated photon (the transition from the level $l = 2$
to $l' = 1$ is chosen as an example) are shown in Fig.~\ref{fig:3} and can
be expressed as:
$$\iota=\frac{dI/du}{\iota_0},$$
where $\iota_0=\alpha (B')^3 (mc^2)^2 /4\hbar$ and equals to
$\sim 10^{9}$~erg/s for the field $B'=0.1$ ($\alpha$ is the fine structure
constant).

Let us estimate radiation intensity by the order of magnitude in two cases.
Some of the neutron stars have a surface magnetic field around
$B\sim 10^{12}$~G
($B' = 0.1$). In such field radiation intensity has the order of magnitude
of $\sim 10^9$~erg/s in the no spin-flip processes. In the case of
energetically favorable spin-flip process ($s=1$, $s ' = - 1$) intensity
is lower by a factor of 10. In the other spin-flip process
($s=-1$, $s ' = 1$) intensity is lower
by a factor of 1000. The ratio between intensities for the spin-flip  and
no spin-flip processes is about 5~\%. This result is well known from the number
of works (see, for example \cite{Bagrov}, \cite{RQT}). Intensity decreases
exponentially if the field becomes lower. For example, in the case of white
dwarfs (field strength is $B\sim 10^8$~G and $B' \sim 10^{ - 5}$ )
intensity is about $\sim 10^{ - 3}$~erg/s and the above ratio is
$\sim $10$^{ - 3}$~{\%}.

%------------------------------------------------------------------------------

\subsection{Ultrarelativistic approximation}
In the ultrarelativistic case radiation intensity defined by Eq.
(\ref{eq62})
\begin{equation}
\label{Rad}
\frac{d^2I^{ss'}}{dyd\Psi } = I_0 \frac{9}{8\pi ^2}\frac{y^2F^2}{\left( {2 + z}
\right)^3\left( {2 + z\left( {1 - y} \right)} \right)^2}D^{s s '} ,
\end{equation}
where $z=3hE/m$, $E$ is the initial electron energy,
$y=\omega(2+z)/Ez$, $F=\sqrt{1+\Psi^2}$,
$\Psi=\psi/\psi_c$, $\psi=\pi/2-\theta$.
Factors $D^{ss'}$ are given by Eqs.(\ref{eq63})-(\ref{eq66}).

Angular distribution of radiation intensity is
symmetrical with respect to the orbit plane if photon polarization is linear.
Indeed, when $V = 0$, $Q = - 1$ (perpendicular polarization) Eqs.~(\ref{eq63})
-- (\ref{eq66}) take on the following form:
\begin{equation}
\label{eq100}
D^{ + + } = 2y^2z^2\left( {\sqrt {\frac{a}{b}} FK_{2 / 3} - K_{1 / 3} }
\right)^2,
\end{equation}
\begin{equation}
\label{eq101}
D^{ - - } = 2y^2z^2\left( {\sqrt {\frac{a}{b}} FK_{2 / 3} + K_{1 / 3} }
\right)^2,
\end{equation}
\begin{equation}
\label{eq102}
D^{ - + } = D^{ + - } = 2y^2z^2\Psi ^2K_{1 / 3}^2 .
\end{equation}

Symmetry about the plane $\psi = 0$ takes place since the above expressions
depend only on the square of the angle $\psi ^2$.
One can see that inequality $D^{ - - } > D^{ + + }$ is always true. Thus,
radiation intensity is greater if particles are in a spin-down state. It is
clear because this state is energetically favorable. Intensities of the
spin-flip processes are equal and radiation is absent in the perpendicular to
the field direction ($\psi=0$). Radiation intensity considerably decreases
if electron spin flips.

In the case of parallel polarization, the expressions (\ref{eq63}) --
(\ref{eq66}) have the form
\begin{equation}
\label{eq103}
D^{ + + } = D^{ - - } = 2\Psi ^2aK_{1 / 3}^2 ,
\end{equation}
\begin{equation}
\label{eq104}
D^{ - + } = 2y^2z^2\left( {FK_{2 / 3} - K_{1 / 3} } \right)^2,
\end{equation}
\begin{equation}
\label{eq105}
D^{ + - } = 2y^2z^2\left( {FK_{2 / 3} + K_{1 / 3} } \right)^2.
\end{equation}
As follows from Eq.~(\ref{eq103}), intensities of no spin-flip processes
coincide. They vanish in the perpendicular to magnetic field direction
($\psi = 0$). Intensity of the energetically unfavorable spin-flip process
is minimal in this direction. Symmetry about the plane $\psi=0$
takes place too.

As follows from Eqs.~(\ref{eq100}) -- (\ref{eq105}), in general,
intensity of the spin-flip process (\ref{eq105}) is
comparable with intensity of the most probable one (\ref{eq101}).
Indeed, the ratio between the differential intensities is equal to
the value $D^{+-}/D^{--}$. In the case of large photon frequency
($\omega \rightarrow E$), the condition
$\sqrt {a/b}=(E + E')/\omega  \rightarrow 1$ is true.
Consequently, $D^{+-}/D^{--} \rightarrow 1$. On the other hand, in
the case of low photon frequency ($\omega \ll E$) $\sqrt {a/b} \gg 1$
and we obtain the well-known result $D^{+-} \ll D^{--}$ \cite{Bagrov}.
This effect is significant when $z\gtrsim 1$, since the maximum
of radiation intensity shifts into the region of high frequency as
the parameter $z$ increases. The same result was obtained numerically in
Ref.~\cite{Preece}.

\begin{figure*}
\begin{center}
\resizebox{\textwidth}{!}{%
  \includegraphics{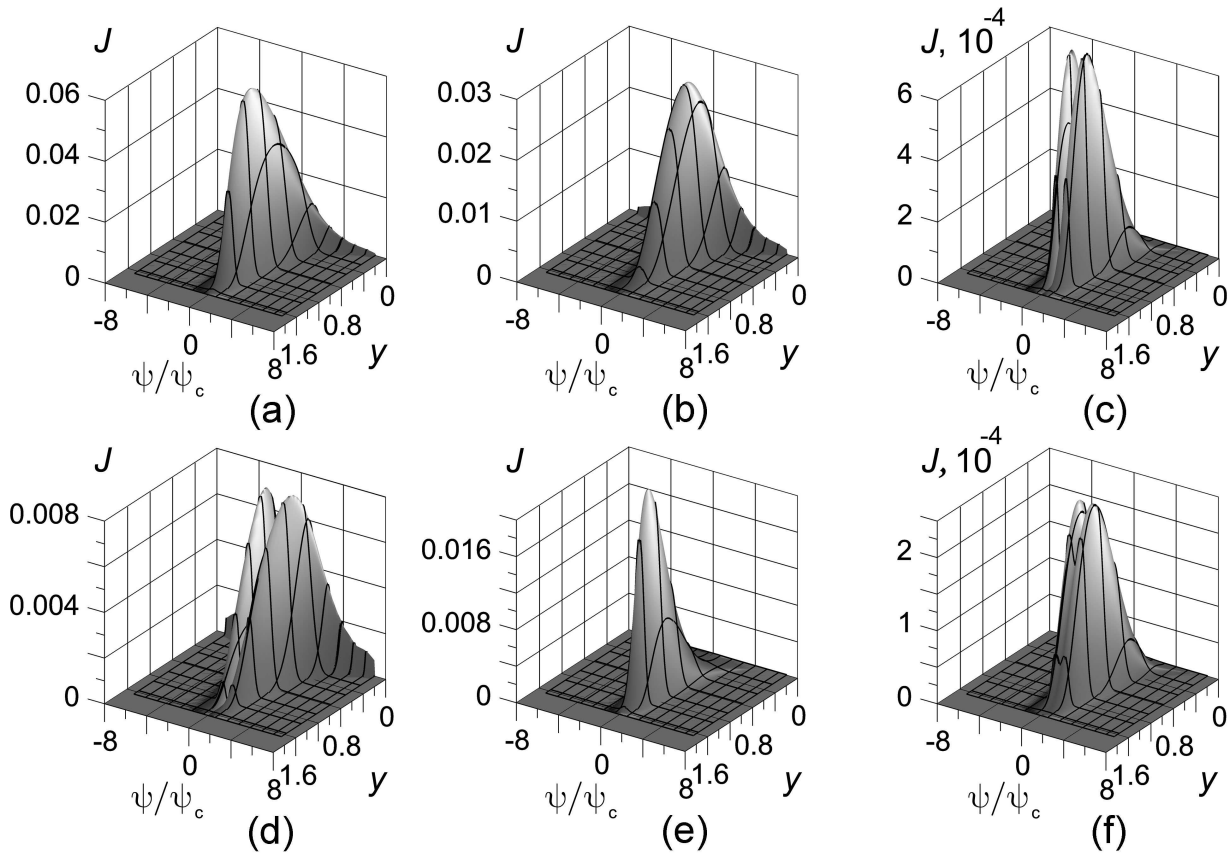}
}
\end{center}
\caption{Dependence of intensity on the output angle and photon frequency
in the case of perpendicular (a)--(c) and parallel (d)--(f) linear polarization
of the photon: (a) $s=s'=-1$; (b) $s=s'=1$; (c)
spin-flip processes; (d) spins of the same orientation; (e) $s=1,s'=-1$;
(f) $s=-1,s'=1$. Here, $J=I/I_0$, $z = 3EB'/m=3$.}
\label{fig:4}
\end{figure*}

The dependence of differential intensity on the output angle and photon
frequency in the case of linear polarization of radiation is shown in
Fig.~\ref{fig:4}.

The following result should be mentioned. It is known that relativistic
particles emit radiation into a narrow cone in the line of motion and
intensity is maximal in the direction of velocity. However, intensity of
perpendicular polarized radiation is zero in the direction $\psi=0$ if flip of
spin occurs. Parallel polarized radiation is absent in this
direction for the no spin-flip processes. Although this effect is
unexpected in the ultrarelativistic approximation, it has general origin.
Indeed, in the LLL approximation, radiation in the line of motion is absent
in the same cases as in the ultrarelativistic approximation.

Angular intensity distribution of circular polarized radiation is not
symmetrical in respect to the plane $\psi=0$. Intensity of radiation of right
circular polarization is maximal in the region $\psi > 0$ and intensity of
the left circular polarized radiation is maximal in the region $\psi < 0$.

In general, the process of synchrotron radiation has similar features in the
ultrarelativistic and the LLL approximations. In the LLL approximation
intensity is maximal along the field direction ($\psi = \pi / 2$) if
polarization is right circular. In the ultrarelativistic approximation the maximum
of right polarized radiation is shifted into the region $\psi > 0$. The shift
of the maximum becomes greater if frequency of the radiated photon decreases.
It is clear since angular distribution of radiation passes to the classical
one in the limit case of small frequency. The situation is reversed if
polarization of radiation is a left circular one.

Note that in the ultrarelativistic approximation intensity depends on the
parameter $z$ only. This parameter is defined by a product of the energy
of the initial electron and the parameter of magnetic field $B'$: $z=3EB'/m$.
Let us estimate intensity of radiation. Let $B'= 10^{ - 5}$ and
$E\sim 50$~GeV. In this case $z = 3$ and radiation intensity per
unit of frequency can be estimated at $\sim 10^{ - 9}$~erg.
%==============================================================================

\section{Spin-polarization effects in the pair production process}
\label{polefpp}
\begin{figure}
\begin{center}
\resizebox{\columnwidth}{!}{
  \includegraphics{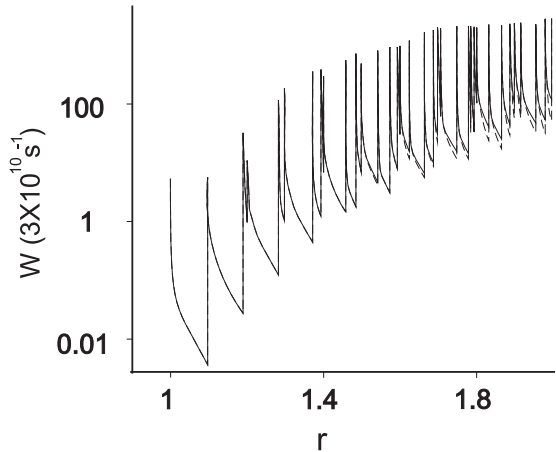}
}
\end{center}
\caption{Dependence of total probability of the pair production process on the
parameter $r = \omega ^2 / 4m^2$. Broken line depicts the result of the
Ref.~\cite{BaierAr}.}
\label{fig:2}
\end{figure}

Probability of the pair production process contains a denominator
that goes to zero if  the pair produced
with zero longitudinal momenta, i.e. at the reaction threshold. It results
in the occurrence of divergences and the process is a resonant one
(Fig.~\ref{fig:2}).

\begin{figure}
\begin{center}
\resizebox{\columnwidth}{!}{
  \includegraphics{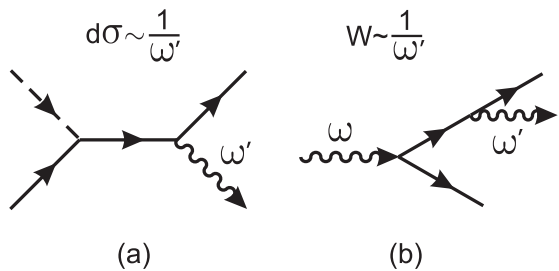}
}
\end{center}
\caption{Feynman diagrams of the processes: (a) bremsstrahlung at the
scattering by a Coulomb center and (b) pair production with radiation of
a final photon. Both probabilities contain the same divergence at
$\omega ' \to 0$}
\label{fig:1}
\end{figure}

Enough attention has been paid to the explanation of the physical nature
of these divergences, for example, in Ref.~ \cite{Graziani}, but there
is not a complete clarity in understanding of this matter.
In our opinion, the presence of singularities is associated with
neglected emission of soft photons, which always accompanies
quantum-electrodynamics processes. This phenomenon is similar to
the so-called ``infra-red catastrophe'' of the
bremsstrahlung process at the scattering by a Coulomb center
\cite{Fomin} (see Fig. \ref{fig:1}).
It is known, that infra-red divergences arise so far as the perturbation
theory becomes incorrect for soft photon emission.

A cross section of the bremsstrahlung process is in inverse proportion to
the frequency of the final photon: $d\sigma \sim 1 / \omega '$. The 
cross section becomes unrestrictedly large if the frequency converges to zero.
Probability of the pair production process has similar dependence on
frequency if an additional final photon is taken into account. In this case,
the divergence at the threshold (longitudinal momentum is zero) vanishes
\cite{Fomin}.

%------------------------------------------------------------------------------
\subsection{LLL approximation}
First of all, obtained probability of pair production
 (\ref{eq81}) -- (\ref{eq84}) does not
depend on the Stokes parameter $V$ that defines  circular polarization.
This fact is a result of the choice of the reference frame where the wave
vector of a photon belongs to the classical orbit plane. Moreover, probability
does not depend on the parameter $U$ too. This parameter defines
polarization in the directions that make angles of $\pm \pi/4$ with the
vector of the magnetic field. However these directions are equivalent since
there is the single preferential direction of vector $\vec B$
in the plane perpendicular to $\vec {k}$.

When an electron and a positron are produced in the low spin
state ($s = - 1$, $s' =1$) the process has the greatest probability because the
corresponding expression (\ref{eq83}) contains the small parameter $B'$
in the lowest power. In the cases $s ' = 1$, $s = 1$ and
$s' = - 1$, $s = - 1$ the expressions of probability (\ref{eq81}),
(\ref{eq82}) differ from Eq. (\ref{eq83}) in the sign of the parameter of
linear polarization $Q$. It should be noted that the similar effect
takes place in the process of synchrotron radiation. If particles are created
in the energetically high spin state ($s = 1$, $s ' = -1$),
the process has the smallest probability.

We may assign an arbitrary value for polarization of the initial photon because
it is defined by the initial conditions of the problem. When
$Q=- 1$ probability of the pair production vanishes in the
energetically low spin state $s = - 1$, $s ' = 1$ (\ref{eq83}).
It is necessary to calculate the probability $W^{ - + }$ in the next order
in small parameter $B'$ before comparison of values of the probabilities.
After the corresponding calculations probability $W^{-+}$
takes on the following form:

\begin{equation}
\label{eq106}
{\begin{array}{l}
 W^{ - +} =\displaystyle \frac{\alpha m^3A}{2\omega }\frac{B'}{p_z g}
 (1 + Q)\times \\
 \times \displaystyle\left[ 1 + \frac{1}{2}B'\left( 3\left( l + l' \right) -
2\frac{ll'}{g^2} \right) \right], \\
 \end{array}}
\end{equation}
where $g = 1 + \left( {p_z /m} \right)^2$. One can see that
dependence of the probability on polarization remains the same as in the
previous case. Thus, the greatest probabilities are
$W^{++}$ and $W^{--}$ in the case of perpendicular polarization ($Q = - 1$).

As follows from above, substantial correlation between polarization
of the initial photon and spin projections of produced particles takes
place. Therefore produced particles are polarized. Let us find the
polarization degree of electrons. By definition, it has the form
\begin{equation}
\label{eq107}
P_ - = \frac{W^+ - W^- }{W^+ + W^- },
\end{equation}
where  $W^ + = W^{ + + } + W^{ + - }$ and $W^ - = W^{ - + } + W^{ - - }$.
The probability $W^{+ - }$ is the smallest one by its order of magnitude
and can be neglected:
\begin{equation}
\label{eq108}
P_ - = \frac{W^{ + + } - W^{ - + } - W^{ - - }}{W^{ + + } + W^{ - + } + W^{
- - }}.
\end{equation}
If $Q \ne - 1$,  the contribution $W^{ - + }$ exceeds all other terms,
therefore  $W^{ + + }$ and $W^{ - - }$ can be neglected. Consequently,
\begin{equation}
\label{eq109}
P_ - \approx - 1.
\end{equation}

Hence, the spins of produced electrons are almost completely oriented against
the field direction if the condition $Q \ne - 1$ is fulfilled.
In order to find the more accurate expression of the polarization degree we
have to substitute Eqs.~(\ref{eq81}), (\ref{eq82}) and (\ref{eq106})
into Eq.(\ref{eq108}) and expand $P_-$ in a power series  in the first
order in small parameter $B'$. After simple calculations, the polarization
degree takes on the form
\begin{equation}
P_ - = - 1 + B'l \frac{1 - Q }{1 + Q }.
\end{equation}

In the case $Q \to - 1$, the quantity $W^{ - + }$ in the expression
(\ref{eq108}) can be neglected and
\begin{equation}
\label{eq110}
P_ - = \frac{l - l'}{l + l'}.
\end{equation}
Consequently, the polarization degree depends on the numbers of Landau levels
of an electron and a positron. The degree of polarization is equal to zero when
the condition $l' = l$ is fulfilled. In the general case the inequality
$\vert P_ - \vert \le 1$ is true. Consequently, produced electrons are
always partially polarized if $l' \ne 0$.

The process has the maximal probability when an electron and a positron are
produced at close or the same Landau levels, therefore polarization
degree converges to zero if the Landau level numbers increase.

Thus, the degree of particle polarization is determined by polarization of
the initial photon. Linear polarization can be changed from a perpendicular one to
a parallel one by rotation of a photon beam by the angle $\pi / 2$ about the
beam axis. It causes substantial changing of the number of particles
in the spin-up and spin-down states. Thus, it is
possible to control the spin orientation of new particles rotating the
photon beam.

Note that the averaged over photon polarization and summed
over particles' spins total probability is in agreement with results of
previous works (Fig.~\ref{fig:2}) \cite{BaierAr}, \cite{Harding1}.
A discrepancy between our result and the computations of Baier and Katkov
is associated with violation of the conditions of LLL approximation.

%------------------------------------------------------------------------------

\subsection{Ultrarelativistic approximation}
In the ultrarelativistic case probability of pair production is given
by Eq.(\ref{eq92}):
\begin{equation}
\label{PP}
W^{s s '} = \int\limits_0^\Omega {\int\limits_{ - \infty }^\infty {W_0
\frac{F^2 D^{s s '} }{24\pi ^2\Omega \varepsilon ^2\left( {\Omega -
\varepsilon } \right)^2}d\varepsilon d\Psi } } ,
\end{equation}
where $\Psi = p_z / E$, $\Omega = B'\omega /m$, $\varepsilon = B'E/m$,
$E$ is the electron energy, and $D^{ss'}$ are defined
by Eqs.(\ref{eq93}) -- (\ref{eq96}).

When the photon is perpendicular polarized the factors
$D^{s s'}$  (\ref{eq93}) -- (\ref{eq96}) have the following forms:
\begin{equation}
\label{eq111}
D^{ - + } = D^{ + - } = 2\Psi ^2K_{1 / 3}^2 ,
\end{equation}
\begin{equation}
\label{eq112}
D^{ + + } = 2\Omega ^2\left( {F\frac{\rho }{\Omega }K_{2 / 3} - K_{1 / 3} }
\right)^2,
\end{equation}
\begin{equation}
\label{eq113}
D^{ - - } = 2\Omega ^2\left( {F\frac{\rho }{\Omega }K_{2 / 3} + K_{1 / 3} }
\right)^2.
\end{equation}

In the case of parallel photon polarization (${Q =1}$), we obtain
\begin{equation}
\label{eq114}
D^{ - + } = 2\Omega ^2\left( {FK_{2 / 3} - K_{1 / 3} } \right)^2,
\end{equation}
\begin{equation}
\label{eq115}
D^{ + - } = 2\Omega ^2\left( {FK_{2 / 3} + K_{1 / 3} } \right)^2,
\end{equation}
\begin{equation}
\label{eq116}
D^{ + + } = D^{ - - } = 2\rho ^2\Psi ^2K_{1 / 3}^2 .
\end{equation}
The expressions (\ref{eq111}) -- (\ref{eq116}) depend on the square of the
angle $\psi^2$ only if polarization of the photon is linear.
Consequently, angular distribution of the probability is symmetrical with
respect to the orbit plane $\psi =0$.

As follows from Eq.(\ref{eq111}), in the case of perpendicular photon
polarization, the probabilities of the processes with opposite particles'
spins are equal to each other. The probabilities $W^{+-}$ and $W^{-+}$ vanish
if the longitudinal momenta of the particles are zero. If the particles have
spins of the same orientation then the corresponding probabilities $W^{++}$
and $W^{--}$ are mirror reflections of each other in the plane $E = E'$.
Indeed, after the replacement $E \leftrightarrows E'$ the argument of the
McDonald functions $X_p = {\omega F^3} / ({3EE'})$ does not change and the
quantity $\rho / \Omega = ( {E - E'} ) / \omega $ changes its sign.

In the case of parallel polarization, the probabilities $W^{++}$ and $W^{--}$
are equal to each other and vanish if the angle $\psi$ goes to zero. These
probabilities also vanish if the energies of the electron and the positron
are equal ($E =E'$), since $\rho = B'\left( {E - E'} \right) / m = 0$ in this
case. One can see from Eqs. (\ref {eq114}), (\ref{eq115}) that the probability
$W^{+-}$ is greater than $W^{-+}$ since the factor $D^{+-}$ is a square of a
sum of nonnegative summands and the factor $D^{-+}$ is a square of a difference
of the same terms. Thus, production of particles in the energetically high
spin state ($s = 1$, $s ' = - 1$)  is more probable than production in
the lower state ($s = - 1$, $s ' = 1$). Note that in the
LLL approximation, the situation is reversed.

It is essential to note that in the cases mentioned above (\ref{eq111})
and (\ref{eq116}), the process is impossible if longitudinal momenta of
particles are zero ($\Psi = p_z / E = 0$). On the contrary,  in the case
of unpolarized particles, probability goes to infinity if the longitudinal
momenta of particles vanish (Fig.~\ref{fig:2}).

\begin{figure*}
\begin{center}
\resizebox{\textwidth}{!}{%
  \includegraphics{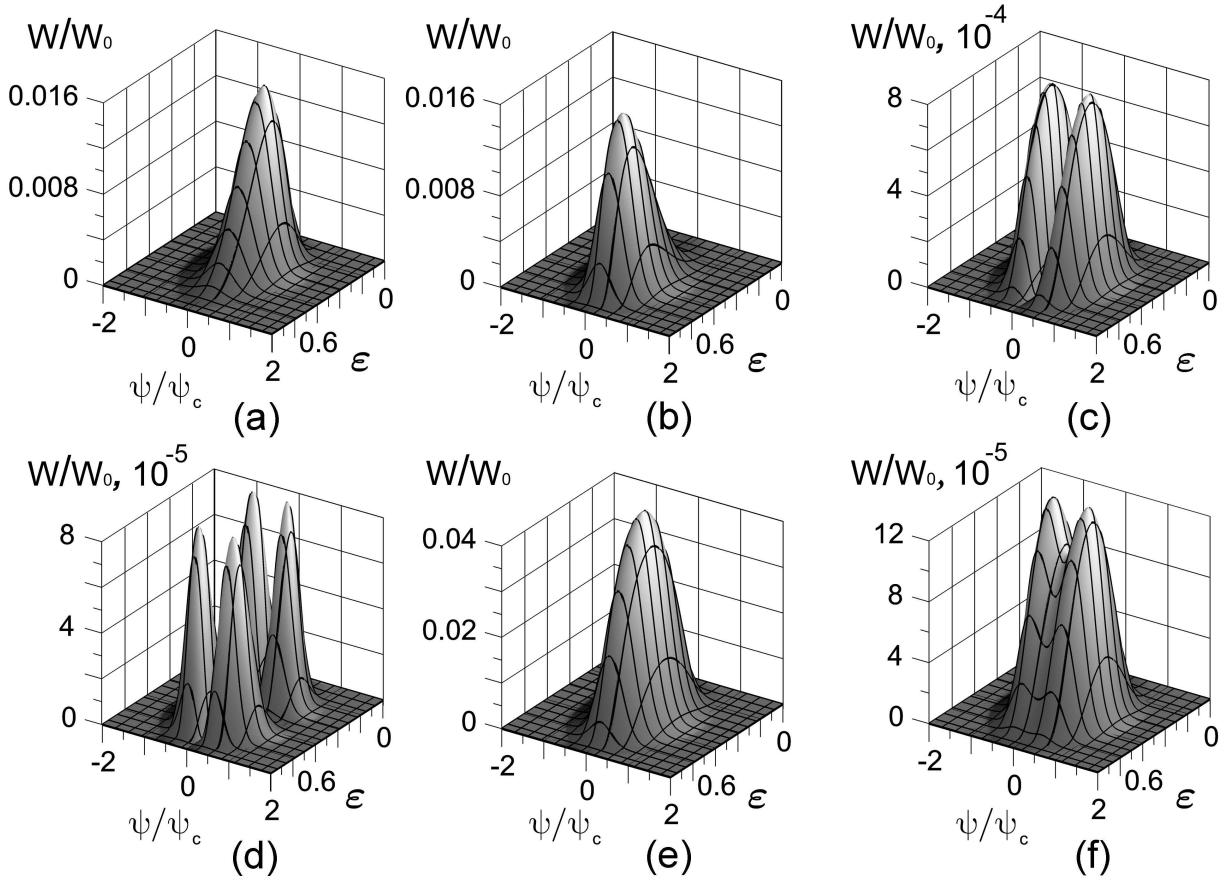}
}
\end{center}
\caption{Dependence of probability of pair production by a photon of
perpendicular polarization (a)--(c) and parallel polarization (d)--(f)
on the electron energy and output angle: (a) $s=s'=1$; (b) $s=s'=-1$;
(c) spins of opposite orientations;
(d) spins of the same direction; (e) $s=1,s'=-1$; (f) $s=-1,s'=1$.
Here, $W_0=\alpha mB'$, $\varepsilon=EB'/m$, $\psi_c=1/\sqrt{2lB'}$,
$\Omega = \omega B'/m = 1$.}
\label{fig:5}
\end{figure*}

Dependence of the process probability on the electron energy and output
angle is shown in Fig.~\ref{fig:5}. Photon polarization is assumed
linear and the value of the parameter $\Omega=\omega B'/m$ is 1.

As follows from the Eqs. (\ref{eq62}) -- (\ref{eq66}), if a photon has circular
polarization, then maximum of probability is shifted with respect
to the plane that is perpendicular to the magnetic field ($\Psi=0$), and cases
of right and left polarizations differ by the shift direction only.

\begin{figure}
\begin{center}
\resizebox{\columnwidth}{!}
{  \includegraphics{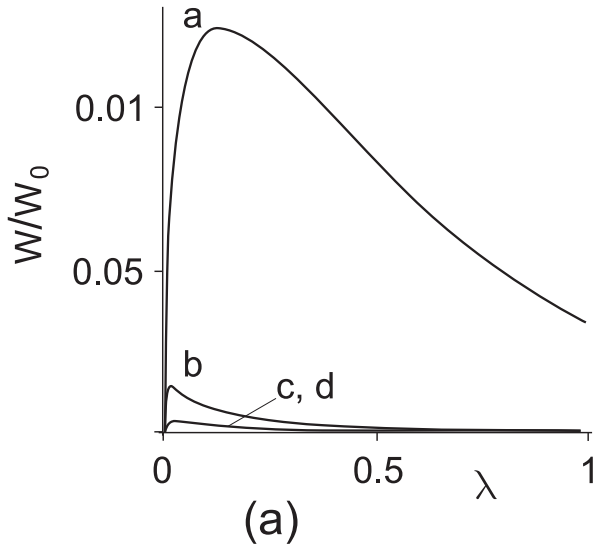} }
\resizebox{\columnwidth}{!}
{  \includegraphics{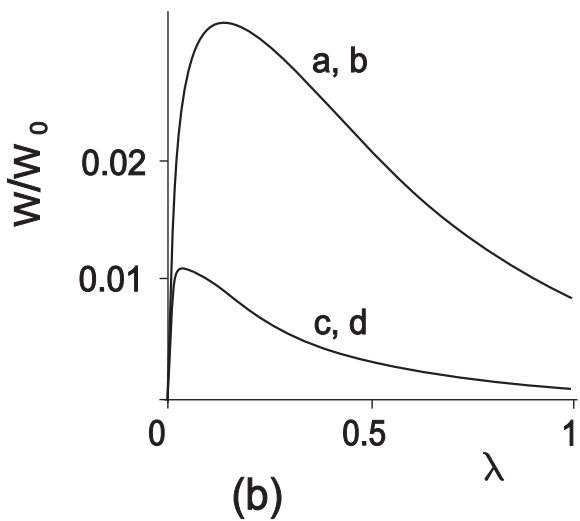} }
\end{center}
\caption{
The dependence of total probability of pair production on the parameter
$\lambda = 4 / 3\Omega $. (a) perpendicular photon polarization; (b) parallel
photon polarization. (a) energetically high spin state of produced particles;
(b) energetically low spin state; (c), (d) spins of the same direction.
}
\label{fig:6}
\end{figure}

Integration of the expressions (\ref{eq93}) -- (\ref{eq96}) over the electron
energy $\varepsilon $ and the output angle $\Psi $ gives the total
probability of the pair production process. The dependence of the total
probability of the process with polarized particles on the parameter
$\lambda = 4 /3\Omega = 4m / 3\omega B'$ is shown in Fig.~\ref{fig:6}. Photon
polarization is assumed to be linear.

One can see that probabilities coincide if particles are produced with
spins of the same orientation. It is clear from the analysis of expressions
(\ref{eq112}), (\ref{eq113}), (\ref{eq116}).

As opposite to the case of the LLL approximation, the process has the largest
probability when a pair is produced in the high spin state ($s = 1$,
$s ' = - 1$) by a photon of parallel polarization ($Q=1$).
In this case, electron spins are almost entirely oriented along the field
and positron spins are oriented against
the field.

In the case of perpendicular polarization, the probabilities also coincide if
particles' spins have
opposite directions. Thus, the beam of produced particles is unpolarized.
As mentioned above, in the LLL approximation, the polarization degree of
electron spins (\ref{eq110}) converge to zero as the numbers of Landau
levels increase. It is in agreement with the obtained result.

Finally, it can be concluded that in the ultrarelativistic approximation
one can control the polarization degree by the setting of the photon
polarization as well as in the LLL approximation.

%==============================================================================
\section{Application}\label{app}

The obtained results can be applied to astrophysical modeling of pulsars.
According to current pulsar models, high energy photons produce
electron-positron pairs in the pulsar magnetic field that subsequently synchrotron
radiate. Particles are considered as unpolarized. However, as follows from Eqs.
(\ref{eq109}), (\ref{eq110}) produced electrons have certain polarization that
is defined by initial photon polarization. Thus, their synchrotron rates are
different from the rates of unpolarized particles. Let us calculate the ratio
$R$ between transition rates of polarized and unpolarized electrons.

\subsection{LLL approximation}

Let $x_+$ be the fraction of spin-up electrons. After corresponding averaging
of Eqs (\ref{eq53})-(\ref{eq56}) the ratio $R$ will take the form
\begin{equation}
R=2\frac{l-x_{+} (l-l')}{l+l'}.
\end{equation}
The fraction $x_+$ can be immediately obtained from Eqs.
(\ref{eq81})-(\ref{eq84}) with the assumption that the electron and the
positron becomes created on the same energy level $l$:
\begin{equation}
x_+=\frac{1}{2}\frac{B'l(1-Q)}{(1+Q)+B'l(1-Q)},
\end{equation}
where $Q$ is the polarization of the initial photon in the pair production
process. One can see that $R=1$ when $x_+=1/2$. The ratio $R$ is greater than
unity if $Q \ne -1$, and probabilities differ twice for ground state
transitions and parallel polarization of the initial photon.

\begin{figure}
\begin{center}
\resizebox{\columnwidth}{!}
{  \includegraphics{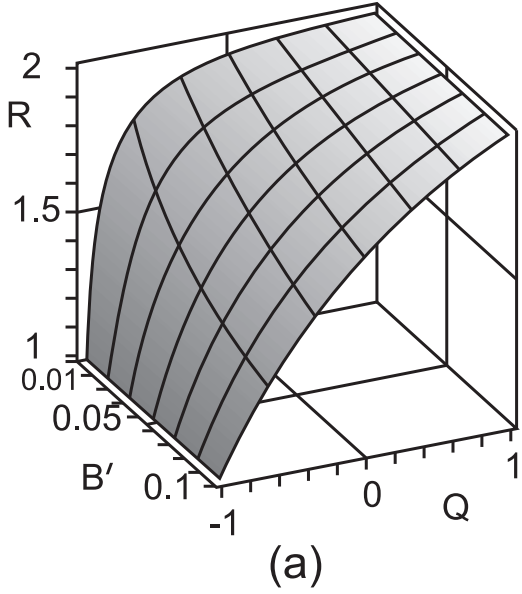} }
\resizebox{\columnwidth}{!}
{  \includegraphics{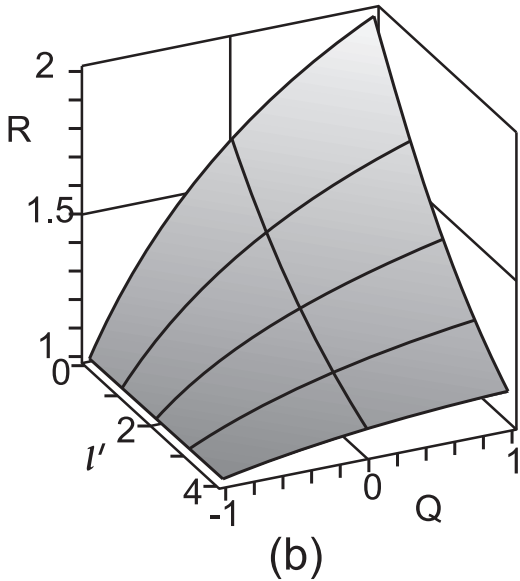} }
\end{center}
\caption{
The dependence of the ratio $R$ between synchrotron rates of polarized
and unpolarized electrons on: (a) photon polarization $Q$ and magnetic
field $B'$, $l=5$, $l'=0$; (b) the number of final Landau level $l'$ and
photon polarization $Q$, $B'=0.1$.
}
\label{fig:7}
\end{figure}

Figure \ref{fig:7}(a) shows the dependence of the ratio $R$ on
photon polarization and magnetic field strength for transition from $l=5$ to
$l'=0$. Figure \ref{fig:7}(b) shows the dependence of the ratio $R$ on
photon polarization and the number of final level. Field strength is $B'=0.1$.

\subsection{Ultrarelativistic approximation}

\begin{figure}
\begin{center}
\resizebox{\columnwidth}{!}
{  \includegraphics{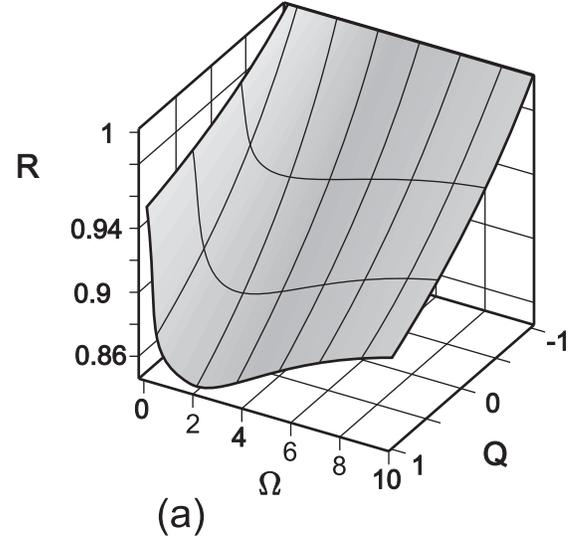} }
\resizebox{\columnwidth}{!}
{  \includegraphics{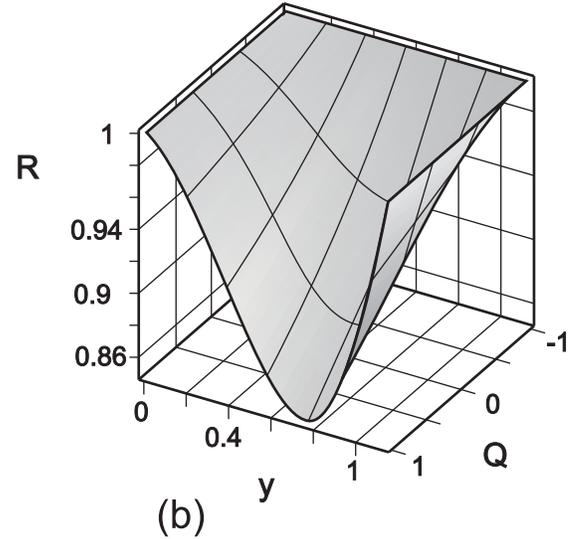} }
\end{center}
\caption{
The dependence of the ratio $R$ between synchrotron rates of polarized
and unpolarized electrons on: (a) initial photon polarization $Q$ and parameter
$\Omega=\omega B'/m$; (b) final photon frequency $y=\omega/\omega_c$ and
initial photon polarization $Q$, $\Omega=10$.
}
\label{fig:8}
\end{figure}

In ultrarelativistic approximation, the ratio $R$ can be obtained by
the integration of Eqs. (\ref{eq62})-(\ref{eq66}) and (\ref{eq92})-(\ref{eq96}).
In Fig.~\ref{fig:8}(a) the dependence of the ratio $R$ on initial photon
polarization $Q$ and parameter $\Omega$ is shown. Initial photon frequency
$\omega=100m$ is adopted and the magnetic field $B'$ changes from value $B'=0.001$
to value $B'=0.1$. Figure~\ref{fig:8}(b) shows the dependence of the ratio $R$
on initial photon polarization $Q$ and final photon frequency
$y=\omega/\omega_c$. The magnetic field is $B'=0.1$.

One can see that $R \leq 1$ in this case. It has minimum value of about
$0.86$ when $Q=1$ and goes to unit if $Q\rightarrow -1$.

Thus, it is substantial to take into account  polarization-dependent spin bias.

%==============================================================================

\section{Conclusions} \label{concl}
In the present work spin and polarization effects in the processes of
synchrotron radiation and electron-positron pair
photoproduction in a strong magnetic field have been considered.
Spin projections and photon polarization have arbitrary values. Obtained
expressions of probability have been analyzed in the ultrarelativistic and
the LLL approximations.

Substantial correlation of spin and polarization effects takes place
in the process of photon radiation.
The dependencies of probabilities of no-flip processes on the Stokes
parameter $Q$ are equal. Probability of spin-up--spin-down
transition contains $Q$ with a reversed sign.

The processes without flip of spin are the most probable in the
LLL approximation. Probability of these processes weakly depends on the polar
angle $\theta$ in the case of perpendicular photon polarization. For parallel
polarization, probability
of radiation equals to zero in the direction $\theta=\pi/2$. In the
ultrarelativistic approximation, the probability of the spin-flip
process $s=1, s'=-1$ is comparable with the probability of
the most probable one if a high energy parallel polarized photon is radiated.

Radiation is absent in the perpendicular to magnetic field
direction in the ultrarelativistic approximation too. Electrons
radiate in a small interval in the vicinity of the orbit plane, but
radiation intensity of perpendicular polarization is equal to zero in the
direction $\theta=\pi/2$ when spin flips. In the case of the processes
without flip of spin, radiation of parallel polarization is absent
in this direction.

In the ultrarelativistic case, maximum of intensity of circular polarized
radiation is shifted relative to the orbit plane. The shift of the maximum
increases if the photon frequency decreases.

Probability of pair production depends on the parameter of linear
polarization $Q$ only. Dependence on the other parameters can be eliminated
by a choice of the reference frame.

Similar to the process of synchrotron radiation, essential spin-polarization
correlation was found. In the LLL approximation, the process has the greatest
probability when the pair is produced in the low spin
state ($s = - 1$, $s' =1$). When spins of produced particles have the same
orientation, process probability contains the Stokes parameter
$Q$ with reversed sign.

In the ultrarelativistic case, probability of particle production with zero
longitudinal momenta ($\Psi = p_z / E = 0$) vanishes if particles have
opposite spins and the photon has perpendicular polarization ($Q=-1$). If
photon has parallel polarization ($Q=1$) then particles with spins of the
same direction cannot be produced with zero longitudinal momenta. In contrast,
in the case of unpolarized particles, probability becomes infinity if
longitudinal momenta go to zero (Fig. \ref{fig:2}).

On the contrary to the LLL approximation, pair production in the high
spin state ($s = 1$, $s'= -1$) by the parallel polarized photon ($Q=1$) has
the largest probability.

Produced particles may have preferred direction of their spins due to the
spin-polarization effects. The degree of particle polarization is determined
by polarization of the initial photon. In both considered approximations,
particles are produced almost completely polarized if photon polarization
is parallel, but the degree of particle polarization converges to zero in
the case of photon of perpendicular polarization ($Q \to - 1$).Thus, it is
possible to control the spin orientation of new particles by rotating the
plane of polarization of the photon beam.

The obtained results are applied to the modeling of pulsars.
Synchrotron rates are compared in two cases: (a) electron spins are equally
populated  and (b) spin populations are determined by polarization of the
initial photon that converts into electron-positron pairs.
Radiation rates coincide for both cases when photon polarization is
perpendicular. In the LLL limit case (a) exceed the other one twice if photon
polarization is parallel. In the ultrarelativistic limit radiation intensity
is greater in case (b) when photon polarization is parallel.

\section*{Acknowledgments}

We thank P. I. Fomin and S. P. Roshchupkin for
their valuable remarks and useful discussions.

%==============================================================================
\appendix
\section{Synchrotron radiation}

We used the same expressions for the wave functions of an electron and a
positron as in Ref.~\cite{FominJETP}. The reference frame where $p_z=0$ is
chosen.

In the LLL approximation, probability of synchrotron radiation is given by
the following expressions (the superscripts denote initial and final
spin projections) \cite{Kholodov}:
\begin{equation}
\label{eq53}
\frac{dW^{ + + }}{du} = \frac{1}{4}\alpha\Lambda ^2B'\omega \left( {1 + u^2 +
2uV - Q \left( {1 - u^2} \right)} \right),
\end{equation}
\begin{equation}
\label{eq54}
\frac{dW^{ - - }}{du} = \frac{1}{4}\alpha\Lambda ^2B'\omega \frac{l}{l'}\left(
{1 + u^2 + 2uV - Q \left( {1 - u^2} \right)} \right),
\end{equation}
\begin{equation}
\label{eq55}
\frac{dW^{ + - }}{du} = \frac{1}{8}\alpha\Lambda ^2(B')^2\omega\frac{\left({l-
l'} \right)^2}{l'}\left( {1 + u^2 + 2uV + Q \left( {1 - u^2}
\right)} \right),
\end{equation}
\begin{equation}
\label{eq56}
\begin{array}{l}
 \displaystyle \frac{dW^{ - + }}{du} =
 \frac{1}{32}\alpha\Lambda ^2(B')^4\omega l\left( {l - l'}
\right)^2\times \\
 \times \left[ {\left( {1 + u^2} \right)\left( {1 + L^2\left( {1 - u^2}
\right)^2} \right) - 2L\left( {1 - u^2} \right)^2 + } \right. \\
 + 2V u\left( {1 - L^2\left( {1 - u^2} \right)^2} \right) + \\
 \left. { + Q \left( {1 - u^2 + L^2\left( {1 - u^2} \right)^3 -
2L\left( {1 - u^4} \right)} \right)} \right], \\
 \end{array}
\end{equation}
Here, $\alpha$ is the fine structure constant, $l$ and $l'$ are the
numbers of initial and final Landau levels, $\omega$ is photon frequency,
$V$ and $Q$ are the Stokes parameters, $L = (l - l')/(l - l' + 1)$,
$u = \cos \theta$ ($\theta$ is a photon polar angle), and
$$\Lambda=e^{ - \frac{\eta }{2}}\eta ^{\frac{l - l' - 1}{2}}\sqrt
{\frac{\left( {l - 1} \right)!}{\left( {l' - 1} \right)!}} \frac{1}{\left(
{l - l' - 1} \right)!},$$
where $\eta={\omega^2 \sin^2 \theta} / {2m^2B'}$.
%------------------------------------------------------------------------------

In the ultrarelativistic approximation intensity of synchrotron radiation
is given by the expression
\begin{equation}
\label{eq62}
\frac{d^2I^{ss'}}{dyd\Psi } = I_0 \frac{9}{8\pi ^2}\frac{y^2F^2}{\left( {2 + z}
\right)^3\left( {2 + z\left( {1 - y} \right)} \right)^2}D^{s s '} ,
\end{equation}
where factors $D^{s s '} $ look like
\begin{equation}
\label{eq63}
{\begin{array}{l}
 D^{ + + }= \left[ \left( \Psi ^2a + b \right)K_{1 / 3}^2 + F^2aK_{2 / 3}^2 -
2FcK_{1 / 3} K_{2 / 3 }\right] + \\
     +  2V\Psi \left[  - cK_{1 / 3}^2 + FaK_{1 / 3} K_{2 / 3 }\right] + \\
     + Q \left[ \left( \Psi ^2a - b \right)K_{1 / 3}^2 - F^2aK_{2 / 3}^2
+ 2FcK_{1 / 3} K_{2 / 3} \right], \\
 \end{array}}
\end{equation}
\begin{equation}
\label{eq64}
{\begin{array}{l}
 D^{ - - }= \left[ \left( \Psi ^2a + b \right)K_{1 / 3}^2 + F^2aK_{2 / 3}^2 +
2FcK_{1 / 3} K_{2 / 3} \right] + \\
     +  2V\Psi \left[ cK_{1 / 3}^2 + FaK_{1 / 3} K_{2 / 3} \right] + \\
     + Q \left[ \left( \Psi ^2a - b \right)K_{1 / 3}^2 - F^2aK_{2 / 3}^2
- 2FcK_{1 / 3} K_{2 / 3} \right], \\
 \end{array}}
\end{equation}
\begin{equation}
\label{eq65}
{\begin{array}{l}
 D^{ - +} = y^2z^2\left\{ \left[ F^2\left( K_{1 / 3}^2 + K_{2 / 3}^2
\right) - 2FK_{1 / 3} K_{2 / 3} \right] + \right.\\
     +  2V\Psi \left[  - K_{1 / 3}^2 + FK_{1 / 3} K_{2 / 3 }\right] + \\
     \left.+ Q \left[ \left( 1 - \Psi ^2 \right)K_{1 / 3}^2 +
     F^2K_{2 /3}^2
 - 2FK_{1 / 3} K_{2 / 3} \right] \right\}, \\
 \end{array}}
\end{equation}
\begin{equation}
\label{eq66}
{\begin{array}{l}
 D^{ + -} = y^2z^2\left\{ \left[ F^2\left( K_{1 / 3}^2 + K_{2 / 3}^2
\right) + 2FK_{1 / 3} K_{2 / 3} \right] + \right. \\
     + 2 V\Psi \left[ K_{1 / 3}^2 + FK_{1 / 3} K_{2 / 3} \right] + \\
    \left.  + Q \left[ \left( 1 - \Psi ^2 \right)K_{1 / 3}^2 +
    F^2K_{2 /3}^2
  + 2FK_{1 / 3} K_{2 / 3} \right] \right\} . \\
 \end{array}}
\end{equation}
Here $K_{1/3}$ and $K_{2/3}$ are the McDonald functions of the argument
$$
X_R = \frac{yF^3}{2 + z\left( {1 - y} \right)}.
$$
The following notations are also used: ${z = 3{EB'}/{m}}$,
$E$ is the initial electron energy,
${y = \omega / \omega _c}$,
${\omega _c ={Ez}/(2 + z)}$,
${I_0 = \alpha(B')^2E^2}$,
${\Psi = \psi / \psi _c}$,
${\psi=\pi/2-\theta}$,
${\psi_c=m/E=1/\sqrt{2lB'}}$ is the characteristic radiation angle,
${F = \sqrt {1 + \Psi ^2}}$,
${a = \left( {4 + z\left( {2 - y} \right)} \right)^2 = \left( {E + E'}
\right)^2 {z^2}/{\omega _c^2 }}$,
${b= y^2z^2 = ( z \omega / \omega _c  )^2}$,
$c = \sqrt {ab}$.

It is possible to carry out integration over angle $\Psi$ \cite{Sokolov1}.
Intensity takes on the form
\begin{equation}
\label{rad1}
I^{ss'}=I_0\int\limits_0^{1+2/z}\frac{3\sqrt{3}}{8\pi}
\frac{yD_y^{ss'}}{(2+z)^3(2+z(1-y))}dy,
\end{equation}
where
\begin{equation}
\label{rad2}
\begin{array}{l}
D_y^{++}=2aK_{2/3}(\lambda)+(b-a)Y(\lambda)-2cK_{1/3}(\lambda)-\\
-Q\left[aK_{2/3}(\lambda)+bY(\lambda)-2cK_{1/3}(\lambda) \right],
\end{array}
\end{equation}
\begin{equation}
\label{rad3}
\begin{array}{l}
D_y^{--}=2aK_{2/3}(\lambda)+(b-a)Y(\lambda)+2cK_{1/3}(\lambda)-\\
-Q\left[aK_{2/3}(\lambda)+bY(\lambda)+2cK_{1/3}(\lambda) \right],
\end{array}
\end{equation}
\begin{equation}
\label{rad4}
\begin{array}{l}
D_y^{-+}=y^2z^2 \left\{2[K_{2/3}(\lambda)-K_{1/3}(\lambda)]+\right.\\
\left.+Q(K_{2/3}(\lambda)-2K_{1/3}(\lambda)+Y(\lambda)) \right\},
\end{array}
\end{equation}
\begin{equation}
\label{rad5}
\begin{array}{l}
D_y^{+-}=y^2z^2 \left\{2[K_{2/3}(\lambda)+K_{1/3}(\lambda)]+\right.\\
\left.+Q(K_{2/3}(\lambda)+2K_{1/3}(\lambda)+Y(\lambda)) \right\},
\end{array}
\end{equation}
\begin{equation}
\label{rad6}
Y(\lambda')=\int\limits_\lambda^\infty K_{1/3}(x)dx, \quad
\lambda=\frac{2y}{2+z(1-y)}
\end{equation}

%==============================================================================
\section{Pair Production}
The reference frame where $\vec{k}\bot\vec{B}$ is chosen.

In the LLL approximation probability of the pair production process
with arbitrary spin projections of particles has the form \cite{Novak}:
\begin{equation}
\label{eq81}
W^{ + + } = \frac{1}{4}\frac{\alpha m^4(B')^2}{\omega E\vert p_z\vert }Al(1-Q),
\end{equation}
\begin{equation}
\label{eq82}
W^{ - - } =\frac{1}{4}\frac{\alpha m^4(B')^2}{\omega E\vert p_z\vert}Al'(1-Q),
\end{equation}
\begin{equation}
\label{eq83}
W^{ - + }= \frac{1}{2}\frac{\alpha m^4B'}{\omega E\vert p_z\vert}A(1 + Q),
\end{equation}

\begin{equation}
\label{eq84}
W^{ + - } = \frac{1}{32}\frac{\alpha m^3(B')^5}{\omega \vert p_z \vert }
All'\left( {({1 + Q }) + \frac{16p_z^2 }{m^2(B')^2}({1 - Q })}
\right);
\end{equation}
where $l$ and $l'$ are the Landau levels of an electron and a positron,
$A$ is a constant that looks like
$$
A = \frac{e^{ - \eta }\eta ^{l + l'}}{l!l'!}.
$$
%

%------------------------------------------------------------------------------

In the ultrarelativistic case pair production probability looks like
\begin{equation}
\label{eq92}
W^{s s '} = \int\limits_0^\Omega {\int\limits_{ - \infty }^\infty {W_0
\frac{F^2 D^{s s '} }{24\pi ^2\Omega \varepsilon ^2\left( {\Omega -
\varepsilon } \right)^2}d\varepsilon d\Psi } } ,
\end{equation}
where
\begin{equation}
\label{eq93}
{\begin{array}{l}
 D^{ - +} = \Omega ^2\left\{ \left[ F^2\left( K_{1 / 3}^2 + K_{2 / 3}^2
\right) - 2FK_{1 / 3} K_{2 / 3} \right] + \right. \\
     + 2 V \Psi \left[ K_{1 / 3}^2 - FK_{1 / 3} K_{2 / 3} \right] + \\
 \left. Q \left[ \left( 1-\Psi^2 \right)K^2_{1/3}+F^2K^2_{2/3}-
 2FK_{1/3}K_{2/3} \right]\right\} ,\\
 \end{array}}
\end{equation}
\begin{equation}
\label{eq94}
{\begin{array}{l}
 D^{ + -} = \Omega ^2\left\{ \left[ F^2\left( K_{1 / 3}^2 + K_{2 / 3}^2
\right) + 2FK_{1 / 3} K_{2 / 3} \right] - \right. \\
     - 2 V\Psi \left[ K_{1 / 3}^2 + FK_{1 / 3} K_{2 / 3} \right] + \\
    \left. + Q \left[ \left( 1 - \Psi ^2 \right)K_{1 / 3}^2 +
    F^2K_{2 /3}^2
 + 2FK_{1 / 3} K_{2 / 3} \right] \right\} , \\
  \end{array}}
\end{equation}
\begin{equation}
\label{eq95}
{\begin{array}{l}
 D^{ + +} = \left[ \left( \rho ^2\Psi ^2 + \Omega ^2 \right)K_{1 / 3}^2 +
F^2\rho ^2K_{2 / 3}^2 - 2F\rho \Omega K_{1 / 3} K_{2 / 3} \right] + \\
     +2 V \Psi \left[ \rho \Omega K_{1 / 3} ^2 -
     F^2\rho ^2K_{1 / 3}K_{2/3}\right] + \\
     + Q \left[ \left( \rho ^2\Psi ^2 - \Omega ^2 \right)K_{1 / 3}^2 -
F^2\rho ^2K_{2 / 3}^2 + 2F\rho \Omega K_{1 / 3} K_{2 / 3} \right], \\
 \end{array}}
\end{equation}
\begin{equation}
\label{eq96}
\begin{array}{l}
 D^{ - - } = \left[ {\left( {\rho ^2\Psi ^2 + \Omega ^2} \right)K_{1 / 3}^2
+ F^2\rho ^2K_{2 / 3}^2 + 2F\rho \Omega K_{1 / 3} K_{2 / 3} } \right] - \\
 -2 V \Psi \left[ {\rho \Omega K_{_{1 / 3} }^2 + F^2\rho ^2K_{1 / 3}
K_{2 / 3} } \right] + \\
 + Q \left[ {\left( {\rho ^2\Psi ^2 - \Omega ^2} \right)K_{1 / 3}^2 -
F^2\rho ^2K_{2 / 3}^2 - 2F\rho \Omega K_{1 / 3} K_{2 / 3} } \right], \\
 \end{array}
\end{equation}
Here $W_0 = \alpha mB'$, $F = \sqrt {1 + \Psi ^2}$,
$\Psi = \psi / \psi _c = p_z / E$, $\Omega = B'\omega /m$,
$\varepsilon = B'E/m$, $\rho = 2\varepsilon - \Omega $.
The argument of McDonald functions $K_{1/3}$ and $K_{2/3}$ is
$$
X_p = \frac{1}{3}\frac{\Omega }{\varepsilon \left( {\Omega - \varepsilon }
\right)}F^3.
$$

After carrying out integration over variable $\Psi$ probability (\ref{eq92})
may be expressed as
\begin{equation}
\label{pp1}
W^{ss'}=\frac{W_0}{8\pi\sqrt{3}}\int\limits_0^\Omega \frac{D_\varepsilon^{ss'}}
{\Omega^2\varepsilon(\Omega-\varepsilon)}d\varepsilon,
\end{equation}
\begin{equation}
\label{pp2}
\begin{array}{l}
D_\varepsilon^{-+}=\Omega^2\left\{2[K_{2/3}(\lambda')-
K_{1/3}(\lambda')]+ \right.\\
\left.+Q[K_{2/3}(\lambda')-2K_{1/3}(\lambda')+Y(\lambda')]\right\},
\end{array}
\end{equation}
\begin{equation}
\label{pp3}
\begin{array}{l}
D_\varepsilon^{+-}=\Omega^2\left\{2[K_{2/3}(\lambda')+K_{1/3}(\lambda')]+
\right.\\
\left.+Q[K_{2/3}(\lambda')+2K_{1/3}(\lambda')+Y(\lambda')]\right\},
\end{array}
\end{equation}
\begin{equation}
\label{pp4}
\begin{array}{l}
D_\varepsilon^{++}=2\rho K_{2/3}(\lambda')-2\rho\Omega K_{1/3}(\lambda')+
(\Omega^2-\rho^2)Y(\lambda')-\\
-Q\left[\rho^2 K_{2/3}(\lambda') -2\rho\Omega K_{1/3}(\lambda') +
\Omega^2Y(\lambda')\right],
\end{array}
\end{equation}
\begin{equation}
\label{pp5}
\begin{array}{l}
D_\varepsilon^{--}=2\rho K_{2/3}(\lambda')+2\rho\Omega K_{1/3}(\lambda')+
(\Omega^2-\rho^2)Y(\lambda')-\\
-Q\left[\rho^2 K_{2/3}(\lambda')+2\rho\Omega K_{1/3}(\lambda') +
\Omega^2Y(\lambda')\right].
\end{array}
\end{equation}
Here, $\lambda'={2\Omega}/{3\varepsilon(\omega-\varepsilon)}$.

It should be noted that total radiation intensity and probability of
photoproduction agree with the results of Refs.~\cite{Klepikov},\cite{Sokolov1}.

%==============================================================================


\begin{thebibliography}{}

\bibitem{Klepikov}
N. P. Klepikov, Zh. Exp. Teor. Fiz. \textbf{26}, 19 (1954).

\bibitem{Tsai1}
W. Tsai, Phys. Rev. D \textbf{8}(10), 3460 (1973).

\bibitem{Tsai2}
W. Tsai and T. Erber, Phys. Rev. D \textbf{10}, 492 (1974).

\bibitem{Tsai3}
W. Tsai, Phys. Rev. D \textbf{10}, 1342 (1974).

\bibitem{Daugherty1}
J. K. Daugherty and I. Lerche, Phys. Rev. D \textbf{14}, 340 (1976).

\bibitem{Daugherty2}
J. K. Daugherty and J. Ventura, Phys. Rev. D \textbf{18}, 1053 (1978).

\bibitem{FIAN}
A. I. Nikishov, in \textit{Trudy FIAN Vol.111} (Nauka, Moscow, 1979),
p. 152.

\bibitem{Sokolov1}
A. A. Sokolov and I. M. Ternov, \textit{Synchrotron radiation}
(Pergamon Press, New York, 1968).

\bibitem{Sokolov2}
A. A. Sokolov and I. M. Ternov, \textit{Synchrotron Radiation
from Relativistic Electrons} (American Inst. of Physics, New York, 1986).

\bibitem{Baier1}
V. N. Baier and V. M. Katkov, Zh. Exp. Teor. Fiz. \textbf{53}, 1478 (1967).

\bibitem{Baier2}
V. N. Baier and V. M. Katkov, Zh. Exp. Teor. Fiz. \textbf{55}, 1542 (1968).

\bibitem{BaierAr}
V. N. Baier and V. M. Katkov, Phys.~Rev.~D \textbf{75}, 073009 (2007).

\bibitem{Herold}
H. Herold, H. Ruder and G. Wunner, Astron. Astrophys.
\textbf{115}, 90 (1982).

\bibitem{Preece}
A. K. Harding and R. Preece, Astrophys. J.
\textbf{319}, 939 (1987).

\bibitem{Pavlov}
G. G. Pavlov, V. G. Bezchastnov, P. Meszaros,
and S. G. Alexander, Astrophys. J. \textbf{380}, 541 (1991).

\bibitem{Schwinger}
J. Schwinger and W. Tsai, Phys. Rev. D \textbf{9}, 1843 (1974).

\bibitem{Orlov}
Yu. F. Orlov and S. A. Kheifets, Pis'ma Zh. Exp. Teor. Fiz. \textbf{2}(8),
513 (1958).

\bibitem{Ternov}
I. M. Ternov, V.G. Bagrov, and R. A. Rzaev, Zh. Exp. Teor. Fiz. \textbf{46}, 374 (1964).

\bibitem{Shabad}
A. E. Shabad, in \textit{Trudy FIAN Vol. 192}, (Nauka, Moscow, 1988), p. 5.

\bibitem{Semionova}
L. Semionova and D. Leahy, Astronomy {\&} Astrophysics \textbf{373},
272 (2001).

\bibitem{Maglab}
N.~Harrison and S.~Croocer, Mag Lab Reports, Vol.~14,Report No 1, p.~11 (2007).

\bibitem{Arzamas}
A. D. Saharov, Usp. Fiz. Nauk, \textbf{161}(5), 29 (1991).

\bibitem{PAST}
P. I. Fomin and R. I. Kholodov, Problems of atomic science and technology
\textbf{6}, 154 (2001).

\bibitem{Darmstadt}
I. Koenig et al. Z. Phys. A
\textbf{346}, 153 (1993).

\bibitem{Harding6}
A. K. Harding, arXiv:astro-ph/0304120v1 7 Apr 2003.

\bibitem{Bussard}
R. W. Bussard, Astrophys. J. \textbf{237}, 970 (1980).

\bibitem{Harding1}
A. K. Harding, Phys. Rep. \textbf{206}(6), 327 (1991).

\bibitem{Harding2}
A. K. Harding and J. K. Daugherty, Astrophys. J. \textbf{374},
687 (1991).

\bibitem{Araya}
R. A. Araya and A. K. Harding, Astrophys. J. Lett.
\textbf{463}, 33 (1996).

\bibitem{Harding3}
A. K. Harding and B. Zhang, Astrophys. J. Lett.
\textbf{548}, 37 (2001).

\bibitem{Ibrahim}
A. I. Ibrahim et al., Astrophys. J. Lett. \textbf{574},
51 (2002).

\bibitem{Harding4}
A. K. Harding, J. V. Stern, J. Dyks, and M. Frackowiak,
Astrophys. J. \textbf{680}, 1378 (2008).

\bibitem{Baring1}
M. G. Baring and A. K. Harding, Astrophys. J. Lett.
\textbf{481}, 85 (1997).

\bibitem{Baring2}
M. G. Baring and A. K. Harding, Astrophys. J. Lett.
\textbf{507}, 55 (1998).

\bibitem{Zhang}
B. Zhang, A. K. Harding, and A. G. Muslimov, The Astrophysical
Journal Letters \textbf{531}, 135 (2000).

\bibitem{Mikheev}
N. V. Mikheev and N. V. Chistyakov, Pis'ma Zh. Exp. Teor. Fiz. \textbf{73}(12), 642 (2001).

\bibitem{Harding5}
A. K. Harding, A. G. Muslimov, and B. Zhang, The Astrophysical
Journal \textbf{576}, 366 (2002).

\bibitem{Thompson}
C. Thompson, Astrophys. J. \textbf{688}, 1258 (2008).

\bibitem{Bagrov}
V. G. Bagrov et al., Zh. Exp. Teor. Fiz.
\textbf{71}, 433 (1976).

\bibitem{RQT}
V. B. Berestetskii, E. M. Lifshitz, and L.P. Pitaevskii, \textit{Relativistic
Quantum Theory} (Pergamon Press, Oxford, 1982).

\bibitem{Graziani}
C. Graziani, A. K. Harding, and R. Sina, Phys. Rev. D \textbf{51}, 7097 (1995).

\bibitem{Fomin}
P.I. Fomin and R.I. Kholodov, Problems of atomic science and technology
\textbf{3}, 179 (2007).

\bibitem{FominJETP}
P.I. Fomin and R.I. Kholodov Zh. Exp. Teor. Fiz. \textbf{117}(2), 319 (2000);
JETP \textbf{90}, 281 (2001).

\bibitem{Kholodov}
R. I. Kholodov and P. V. Baturin, Ukr. J. Phys. \textbf{46}(5-6), 621
(2001).

\bibitem{Novak}
A. P. Novak and R. I. Kholodov, Ukr. J. Phys. \textbf{53}(2), 185 (2008).

\end{thebibliography}
\end{document}